\documentclass[showpacs,pra,twocolumn,aps]{revtex4}
\usepackage[T1]{fontenc}
\usepackage[latin1]{inputenc}
\usepackage{bm}
\usepackage{amsmath}
\usepackage{amssymb}
\usepackage{latexsym}
\usepackage{amsfonts}
\usepackage{epsfig}
\usepackage{psfrag}
\newcommand{\ket}[1]{\left|#1\right>}
\newcommand{\bra}[1]{\left<#1\right|}

\newcommand{\nn}{\nonumber\\}

\newcommand{\bea}{\begin{eqnarray}}
\newcommand{\ea}{\end{eqnarray}}
\newcommand{\eea}{\end{eqnarray}}
\newcommand{\ord}{{\cal O}}
\newcommand{\sumint}[1]

\begin{document}

\title{Bogoliubov theory of quantum correlations in the time-dependent
  \\ Bose-Hubbard model}   

\author{Uwe R. Fischer$^{1}$, 
Ralf Sch\"utzhold$^{2}$ and Michael Uhlmann$^{2,3}$} 

\affiliation{$^{1}$Eberhard-Karls-Universit\"at T\"ubingen,
Institut f\"ur Theoretische Physik\\
Auf der Morgenstelle 14, D-72076 T\"ubingen, Germany
\\
$^{2}$Institut f\"ur Theoretische Physik,
Technische Universit\"at Dresden, D-01062 Dresden, Germany \\
$^3$Department of Physics, Australian National University, Canberra,
ACT 0200, Australia} 
\begin{abstract}
By means of an adapted mean-field expansion for large fillings
$n\gg1$, we study the evolution of quantum fluctuations in the
time-dependent Bose-Hubbard model, starting in the superfluid state
and approaching the Mott phase by decreasing the tunneling rate or 
increasing the interaction strength in time.
For experimentally relevant cases, we derive analytical results
for the temporal behavior of the number and phase fluctuations, 
respectively.  
This allows us to calculate the growth of the quantum depletion and
the decay of off-diagonal long-range order.  We estimate 
the conditions for the observability of the time dependence in the 
correlation 
functions in the experimental setups with external trapping present.  
Finally, we discuss the analogy to quantum effects in the early
universe during the inflationary epoch.  
\end{abstract}

\pacs{
73.43.Nq, 
03.75.Lm, 
03.75.Kk, 
05.70.Fh. 
}
 
\maketitle

\section{Introduction}

The rapidly growing interest in the exploration of the dynamics of
quantum phase transitions \cite{SachdevBook} fosters our understanding
of the complex behavior of many-body systems far from equilibrium. 
This concerns, in particular, the interplay of the microscopic degrees
of freedom, their entanglement, and the resulting emergent behavior. 
The controlled study of the time development of correlation functions
expressing these entanglement properties, and the dynamical emergence
of correlations from an initially uncorrelated state has been
undertaken for various systems.  
The Bose-Hubbard model, describing the essential archetype of an
emergence of strong correlations when one crosses a quantum phase
transition point, was originally introduced in a conventional
condensed matter context, to explain certain properties of
bosons in periodic and/or random potentials \cite{BoseHubbard}.  
Its implementation with ultracold atoms \cite{BosonicMott} and the 
subsequent experimental realization 
of the superfluid-Mott transition 
\cite{Greiner} has caused a flurry of research activity. 

This activity is reviewed from a theoretical point of view in
\cite{Lewenstein}, while a number of recent experimental efforts
studying ultracold atoms in optical lattices are covered, e.g., in
\cite{Morsch}.  
Initially, theoretical studies were primarily 
devoted to the transition from the Mott
regime, where number fluctuations are frozen (for commensurate filling
of the lattice sites), to the superfluid side of the transition for
which, conversely, phase fluctuations are frozen 
\cite{Polkovnikov,ClarkJaksch,Sengupta,Isella,Cucchietti}.   
Most of these investigations were done numerically, with the exception
of certain exactly solvable cases like the Ising chain in a transverse 
field \cite{Dziarmaga}.

We discuss here in detail a mean-field approach, first
presented in \cite{BH}, which enables the (in some particular cases
analytical) rigorously controlled calculation of quantum correlations
developing in rapid quenches from the superfluid to the Mott
insulating phase.   
Such a number-conserving and hence controlled mean-field approach is
valid at large filling $n\gg1$ of the lattice sites, with the square
root of the inverse filling $1/\sqrt{n}$ providing the expansion
parameter \cite{Kollath}.  
Here, we supply in particular analytical estimates for the
applicability of the results obtained in \cite{BH} to the
experimentally relevant harmonically trapped case, by comparing the
decay time of superfluid coherence to the propagation of the
disturbances in the system induced by the quench. 
It should be noted that the mean-field ``hydrodynamic'' 
limit of large site fillings considered in the following 
is not of purely academic interest. 
While many experiments on the Mott transition are carried out at small 
filling of order unity, experiments on number squeezing at large
filling have indeed been performed as well
\cite{Orzel,Tuchman,Li}, already on an early stage of research into
the possible occurrence of the Mott insulator transition \cite{Orzel}.   

\section{The Bose-Hubbard Model at Large Filling} 

The Bose-Hubbard model describes interacting bosons on a lattice,
hopping from site to site. 
Restricting ourselves to (two-body) {contact} interactions, i.e., that
two bosons interact only if at the same site $\alpha$, and to one 
single-particle state at each site, adding a
one-particle scalar potential term, the Hamiltonian reads 
\bea
\label{Hamiltonian}
\hat H
=
J \sum\limits_{\alpha\beta}
M_{\alpha\beta}\hat a_\alpha^\dagger\hat a_\beta
+\frac{U}{2}\sum\limits_\alpha(\hat a_\alpha^\dagger)^2\hat a_\alpha^2
+ \sum\limits_\alpha V_\alpha \hat n_\alpha
\,. \label{Hubbard}
\ea
The coupling constant $U$ is linear in the bulk contact interaction 
coupling constant $g$. 
The externally imposed scalar potential to additionally confine the atoms in 
the lattice is in most experimental situations to date to a good
approximation harmonic, $V_\alpha \propto \alpha^2$, 
or linear like in \cite{Li}, $V_\alpha \propto \alpha$.  
The matrix $M_{\alpha\beta}$ describes the fact that, in the lattice,
the effective mass (and possible higher terms in a gradient
expansion), can in general depend on position and the direction of
hopping of the particles from site to site. 
For the simplest example of a one-dimensional chain with
nearest-neighbor hopping, we have 
$M_{\alpha\beta}=\delta_{\alpha,\beta}-\frac14
(\delta_{\alpha,\beta+1}+\delta_{\alpha,\beta-1}+
\delta_{\alpha-1,\beta} + \delta_{\alpha +1,\beta} )$, and the 
effective mass of the bosons 
is given by $1/m^* = J a^2$, where $a$ is the lattice spacing.

At large (average) filling $n\gg1$ the Bose-Hubbard Hamiltonian
\eqref{Hubbard} can be mapped to the so-called quantum rotor model 
(cf., e.g., \cite{SachdevBook,GarciaRipoll}). 
Insertion of the (quantum) Madelung transformation 
$\hat a_\alpha = e^{i\hat \phi_\alpha}\sqrt{\hat n_\alpha}$ 
into \eqref{Hubbard} and expansion into inverse powers of $n$ yields
the quantum rotor Hamiltonian 
\bea
\hat H &=& \sum_\alpha \left\{ 
- n J \cos [\hat \phi_\alpha -\hat \phi_{\alpha+1}]
+\frac U2 (\hat n_\alpha-n)^2 \right\}  \nn 
&=& \sum_\alpha \left\{ - n J \cos [\hat \phi_\alpha -\hat \phi_{\alpha+1}]
-\frac U2 \frac{\partial^2}{\partial \hat \phi_\alpha^2} \right\},  
\label{rotorH} 
\ea
where we have taken for simplicity a 1D lattice with no external
trapping, $V_\alpha=0$. 
As an important ingredient, we use in this representation that local
number fluctuation $\delta \hat n_\alpha = \hat n_\alpha-n$ and phase
variables are conjugate in the limit of large $n$, 
\bea
\left[ \hat\phi_\alpha,\delta \hat n_\beta \right]
=
i\delta_{\alpha\beta}
\,. 
\label{conjugate} 
\ea 
The fact that the number and phase variables are canonically conjugate
variables (implying the very existence of a  phase operator) depends
on applying the mean-field (and effectively hydrodynamic,
i.e. coarse-grained) limit $n\rightarrow \infty$ has been known for a
long time cf., e.g., \cite{Froehlich,CastinII}. 
The problems arising without the limit $n\rightarrow \infty$ can be
seen by means of the full commutator 
\bea
\left[ \hat\phi_\alpha,\hat n_\beta \right]
\stackrel{?}{=}
i\delta_{\alpha\beta}
\,. 
\label{full} 
\ea 
Since $\hat n_\beta$ possesses a discrete spectrum with proper
eigenvectors $\ket{n_\alpha}$, taking the $\ket{n_\alpha}$-expectation 
value of the above relation yields for $\alpha=\beta$ a contradiction. 
In the large $n$-limit, however, the normalized fluctuations 
$\delta\hat n_\alpha/n$ have quasi-continuous spectra and in this
sense Eq.~\eqref{conjugate} provides a valid effective description. 

\begin{figure}[hbt]
\begin{center}
\mbox{\epsfxsize=4cm\epsffile{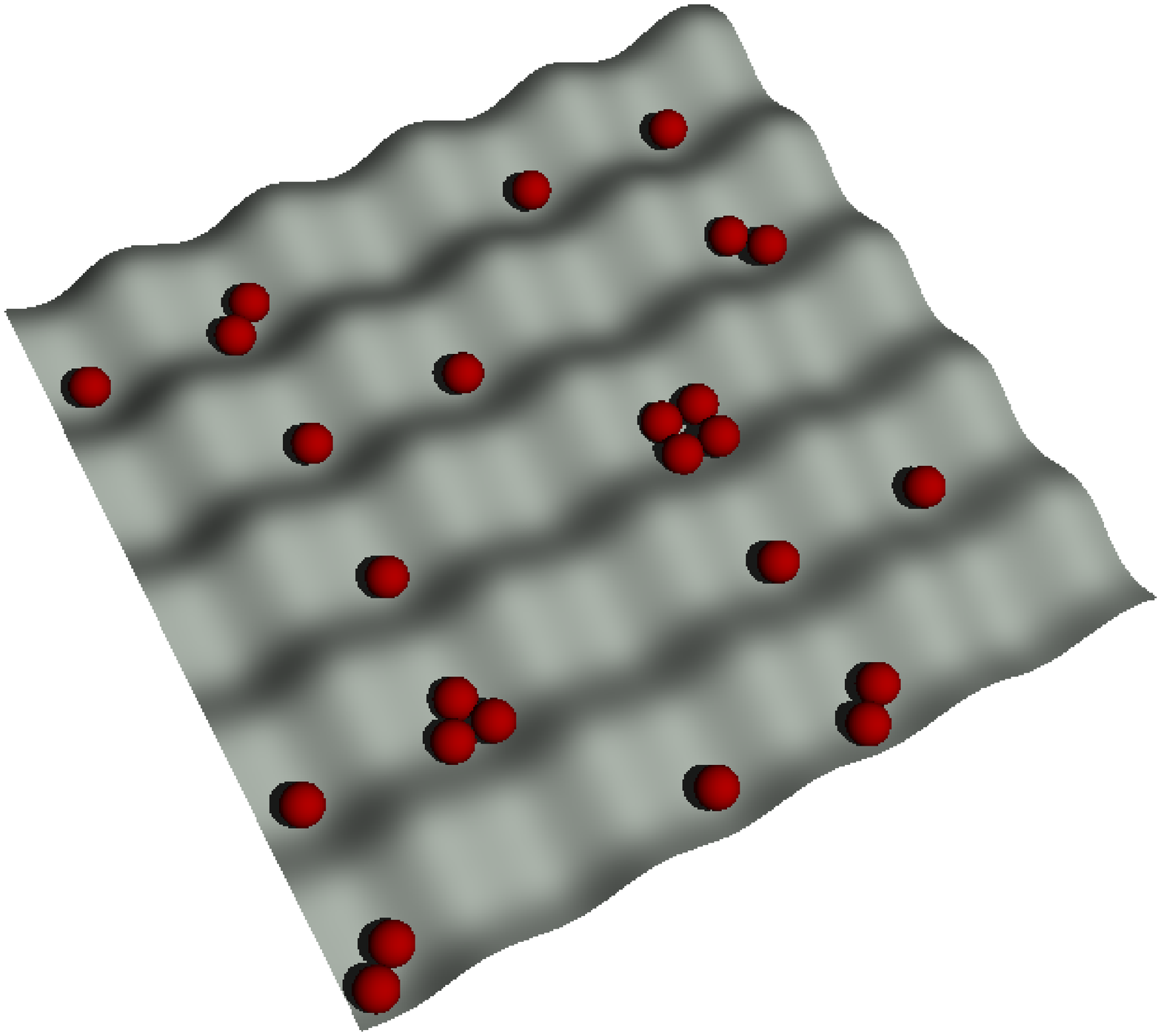}}
\hspace{.2cm}
\mbox{\epsfxsize=4cm\epsffile{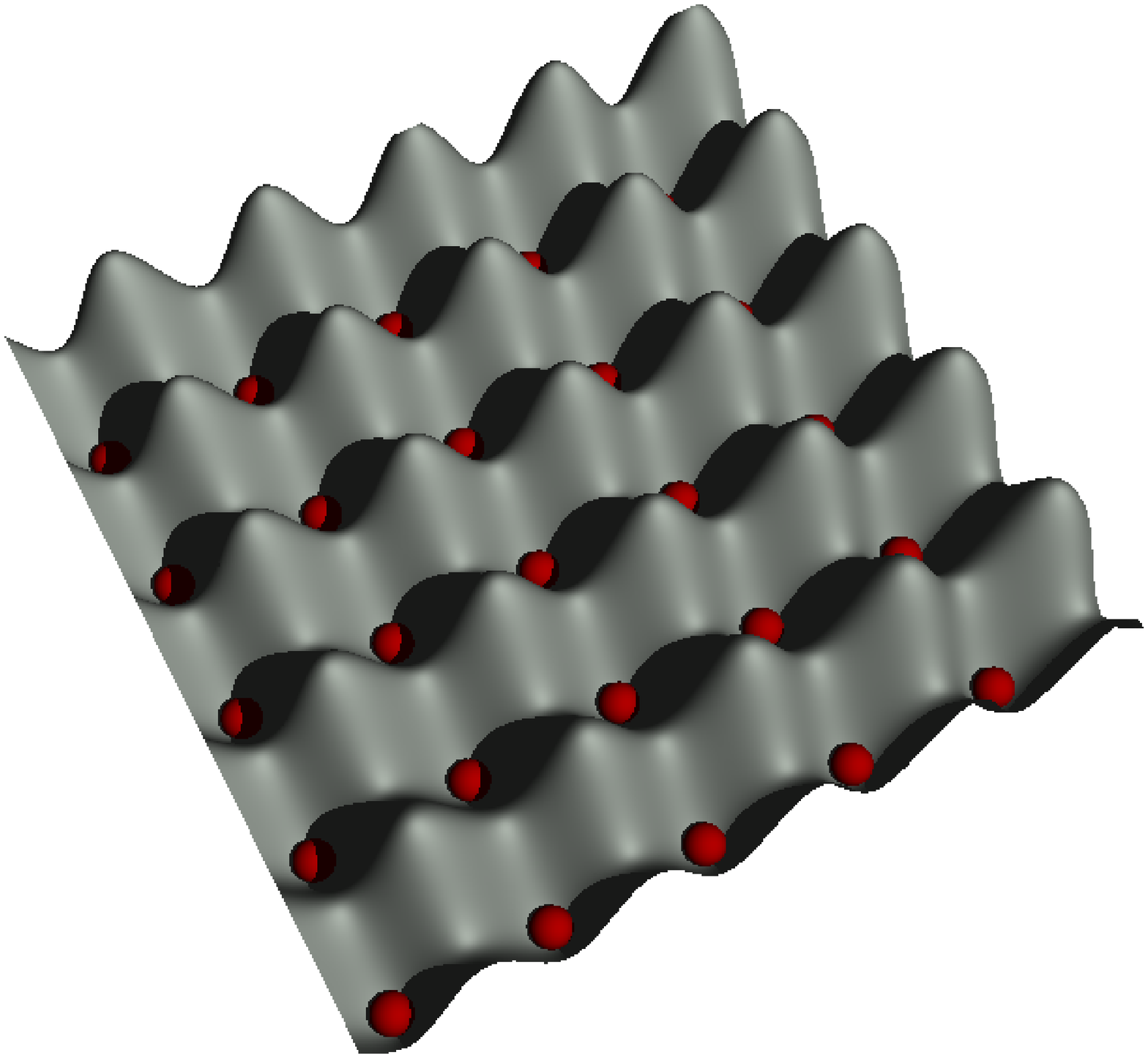}}
\caption{
\label{MottTrans} [Color online] Localization (superfluid--Mott) 
transition of particles in an optical lattice.
For increasing lattice depth and commensurate filling (integer number of
particles per site), the system performs a quantum phase transition 
from a superfluid declocalized phase with highly mobile bosons (left) to 
a localized insulator phase with immobile bosons at commensurate filling 
(right).
The present analysis investigates the temporal evolution of correlations
when one is rapidly going from the left (superfluid) to the right (Mott) side. 
For the sake of this representation, 
we display the non-mean-field case of unit filling, $n=1$, here.
}
\end{center}
\end{figure}

The form \eqref{rotorH} provides a nice and intuitive understanding of
the two phases:
In the superfluid state, the first term on the right-hand side dominates and 
$\hat \phi_\alpha -\hat \phi_{\alpha+1}$ is small (phase coherence). 
In the Mott phase, on the other hand, the second term wins and the
number fluctuations $\delta \hat n_\alpha$ become small.
Therefore, we may directly read off the scaling of the critical point
$J_c$, which separates the two regimes $J_c = \ord(U/n)$. 
We depict a schematic representation of the superfluid-Mott transition 
on a quadratic 2D lattice in Fig.\,\ref{MottTrans}.  

\section{Dynamical mean-field theory} 

We first consider the case without external trapping, $V_\alpha=0$,  
for which the filling is homogeneous $\langle\hat n_\alpha\rangle=n$. 
From \eqref{Hubbard},  
the Heisenberg equations of motion for the lattice field operator are 
(we take $\hbar \equiv 1$ throughout)
\bea 
\label{full-eom}
i\partial_t\hat a_\alpha=
J\sum_{\beta}M_{\alpha\beta}\hat a_\beta+
U\hat n_\alpha\hat a_\alpha 
\,.
\ea
To proceed, we now employ a lattice version of the number-conserving
mean-field expansion which in its continuum version has been introduced in
\cite{Particle,Castin}.
Since the Hamiltonian \eqref{Hubbard} cannot be diagonalized exactly,
a controlled analytical approach requires a small or large parameter
such that we can employ an expansion into powers of this control
parameter. 
One option would be the weak-coupling regime $J\gg U$, where an
expansion into powers of $U/J$ justifies the mean-field
approach.
However, since we want to study the sweep to the Mott phase, this
ratio does not remain small and hence cannot be used as a control
parameter. 
Therefore, we focus on the case of large fillings $n\gg1$ here and use
$1/n$ as a small expansion parameter \cite{supersolid}. 
I.e., in the limit of large fillings $n$, the full site field operator
is expanded into terms of various power in $n$,  
\bea
\label{mean-field}
\hat a_\alpha
=\left(\psi_0+\hat\chi_\alpha+\hat\zeta_\alpha\right)
\frac{\hat A}{\sqrt{\hat N}}
\,. 
\ea
The operator
$\hat A=\hat a_\Sigma(\hat a_\Sigma^\dagger\hat a_\Sigma)^{-1/2}\hat N^{1/2}$
with $\hat a_\Sigma=\sum_\alpha\hat a_\alpha$ 
accounts for the conservation of the total number of particles 
$\hat N=\hat A^\dagger\hat A=\sum_\alpha\hat n_\alpha$. 
Due to the number-conserving nature of the ansatz, 
we have $\langle\hat a_\alpha\rangle=0$ exactly, as opposed to 
non-number-conserving approaches, in which 
$\langle\hat a_\alpha\rangle=\ord(\sqrt n)$. We stress that 
a number-conserving formalism is essential to describe correlation 
functions accurately, especially those of higher order.

As stated before, the above mean-field expansion requires that, at
each site, the integral filling 
$n_\alpha=\langle\hat n_\alpha \rangle = 
\langle\hat a_\alpha^\dagger\hat a_\alpha\rangle\in\mathbb N$ 
is much larger than unity, $n_\alpha\gg 1$.
The idea of \eqref{mean-field} is to expand the original operator 
$\hat a_\alpha$ into powers of $n_\alpha\gg1$, considering the
(formal) limit in which $n \uparrow\infty$ but with the chemical
potential $\mu=Un$ remaining finite and fixed, such that $U=\ord(1/n)$.
The leading term in the above expansion \eqref{mean-field} 
is the order parameter $\psi_0=\ord(\sqrt n)$ describing the
condensate part. 
The linearized quantum corrections $\hat\chi_\alpha=\ord(n^0)$ are
decomposed into single-particle contributions, and correspond to
Bogoliubov quasiparticle excitations above the superfluid ground state,  
after a Bogoliubov transformation to a quasiparticle basis.
For the validity of the expansion \eqref{mean-field}, the remaining
(non-linear) higher-order corrections $\hat\zeta_\alpha$ containing
multi-particle contributions must remain small,
$\hat\zeta_\alpha\ll1$, 
during the whole temporal evolution of the system. 
Since we are dealing with formally unbounded operators, the condition 
$\hat\zeta_\alpha\ll1$ should be understood as a weak limit, i.e., 
$\langle\hat\zeta_\alpha\rangle\ll1$ and 
$\langle\hat\zeta_\alpha^\dagger\hat\zeta_\alpha\rangle\ll1$ etc. 

Thus, $\hat\zeta_\alpha$ of order unity signifies that one approaches
strongly correlated states, i.e., the Mott regime. 
These ``strong correlations'' are due to interactions between the
quasi-particle excitations $\hat\chi_\alpha$, which become stronger
when approaching the transition, and eventually generate a gap due
to non-perturbative effects.
The emergence of this gap denotes the transition from the delocalized
atoms (superfluid state) to the localized atoms (Mott state), 
see, e.g., \cite{BDZReview}.  
In what follows, we are interested in describing the approach to the  
localized state of the atoms, and how in a dynamical quantum phase 
transition the corresponding decrease in the phase ordering takes place.   

The Bogoliubov-de~Gennes equations for the quasiparticle excitations
at site $\alpha$, which  are by definition described by the first
order operators $\hat\chi_\alpha$, read 
\bea
\label{Bogoliubov-deGennes}
i\partial_t\hat\chi_\alpha
=
J\sum\limits_{\beta}M_{\alpha\beta}\hat\chi_\beta
+2U|\psi_0^2|\hat\chi_\alpha+U\psi_0^2\hat\chi_\alpha^\dagger
\,,
\ea
with $\psi_0$ being determined by the solution of the 
Gross-Pitaevski\v\i\/ mean-field equation 
\bea
\label{GP}
i\partial_t\psi_0=
J\sum_{\beta}M_{\alpha\beta}\psi_0+
U|\psi_0|^2\psi_0 
=
U|\psi_0|^2\psi_0
\,, 
\ea
where the final equality holds for a spatially homogeneous ground state. 
All the residual terms, remaining after the insertion of
\eqref{mean-field} into \eqref{full-eom}, determine the higher-order
corrections $\hat\zeta_\alpha$ to the mean-field expansion
\eqref{mean-field}, which then evolve according to the equation of
motion  
\bea
\label{higher-order}
i\partial_t\hat\zeta_\alpha
&=&
J\sum\limits_{\beta}M_{\alpha\beta}\hat\zeta_\beta
+2U|\psi_0^2|\hat\zeta_\alpha+U\psi_0^2\hat\zeta_\alpha^\dagger+
\\
&&
+2U\psi_0\hat\chi_\alpha^\dagger\hat\chi_\alpha
+U\psi_0^*\hat\chi_\alpha^2+U\hat\chi_\alpha^\dagger\hat\chi_\alpha^2
+\ord(U\hat\zeta_\alpha)
\nonumber
\,.
\ea
Deep in the superfluid phase (our initial state), the higher-order 
corrections $\hat\zeta_\alpha$ are small and the mean-field
expansion~(\ref{mean-field}) works very well.
If we approach the Mott phase, however, these corrections start to 
grow according to Eq.\,(\ref{higher-order}) and, at some point, the
mean-field expansion~(\ref{mean-field}) breaks down.
The characteristic time-scale of this breakdown can be estimated from
the nonlinear source terms in Eq.\,(\ref{higher-order}), which are
suppressed to $\ord(1/\sqrt{n})$ in view of the imposed 
constancy of the chemical potential, so that
$U=\ord(1/n)$, $\psi_0=\ord(\sqrt{n})$, and
$\hat\chi_\alpha=\ord(n^0)$. 
Therefore (starting in the superfluid phase), the higher-order
corrections remain small as long as we have ${Ut\sqrt{n}\ll 1}$, i.e., 
for evolution times which are of order $t=\ord(\sqrt{n})$. 
Thus, while Bogoliubov mean-field theory is, in principle, also valid 
(initially, i.e., for $J\gg U$) for small values of $n=\ord(1)$, the
$\hat\zeta_\alpha$ corrections in \eqref{higher-order}, which describe
correlations beyond those derivable from the mean-field plus
single-particle fluctuations, i.e., from $\psi_0$ and $\hat
\chi_\alpha$ in \eqref{mean-field}, begin to grow very quickly in
off-equilibrium situations, then invalidating the Bogoliubov approach
at small filling. 
For large fillings $n\gg1$, however, the mean-field expansion
\eqref{mean-field} can be extrapolated to relatively long time-scales
$t=\ord(\sqrt{n})$, which allows us to study the sweep analytically
(with $\hat\zeta_\alpha$ serving as an indicator for the validity of
the mean-field approach). 

Linearizing the polar decomposition of the fundamental field operator 
$\hat a_\alpha=\exp\{i\hat\phi_\alpha\}\sqrt{\hat n_\alpha}$, we may
identify the fluctuations 
$\hat\chi_\alpha= \psi_0[\delta\hat n_\alpha/(2n)+i\delta\hat\phi_\alpha]
+\ord(1/\sqrt{n})$ in terms of the linearized number fluctuations
$\delta\hat n_\alpha$ and the conjugate phase fluctuations
$\delta\hat\phi_\alpha$ according to \eqref{conjugate}. 
Eq.\,(\ref{Bogoliubov-deGennes}) can be diagonalized by a
normal-mode expansion into the eigenvectors 
$M_{\alpha\beta} v^\beta_\kappa = \lambda_\kappa v^\alpha_\kappa$ 
of the hopping matrix $M_{\alpha\beta}$,  
with the eigenvalues $\lambda_\kappa$ 
labeled by the generalized momenta $\kappa$. 
To this end, we expand the number and phase fluctuations via 
$\delta\hat n_\kappa = v^\alpha_\kappa \delta \hat n_\alpha$
and 
$\delta\hat\phi_\kappa = v^\alpha_\kappa \delta \hat\phi_\alpha$,
respectively, and insert them into 
\bea 
\hat\chi_\alpha = \sqrt n\, e^{-i n \int U dt} (\delta \hat n_\alpha/2n 
+i\delta\hat\phi_\alpha), 
\ea
where we have factored out the time-dependence of the
Gross-Pitaevski\v\i\/ wave function with the appropriate 
phase factor $\psi_0 = \sqrt n  e^{-i n \int U dt}$
as follows from \eqref{GP}. 
Thus, we obtain from the Bogoliubov-de~Gennes equations 
\eqref{Bogoliubov-deGennes} two real equations for the time evolution 
of density and phase fluctuations in the eigenbasis labeled by $\kappa$, 
\bea 
\partial_t \delta \hat n_\kappa  &=& 
2n J \lambda_\kappa \delta\hat\phi_\kappa\,, \nn
\partial_t \delta \hat \phi_\kappa  &=& -\left(\frac{J \lambda_\kappa}{2n}
+U\right)\delta\hat n_\kappa 
\,, 
\label{dotdelta}
\ea
for generally time-dependent $J$ and $U$.  
In case that both $J$ and $U$ are constant in time, the above equations
result in the well-known Bogoliubov spectrum \cite{Bogoliubov,BdGfootnote} 
\bea
\omega_\kappa^2 = J^2 \lambda_\kappa^2 +2n U J \lambda_\kappa\,.
\label{spectrum} 
\ea 

We now discuss two possible routes to approach or cross the phase
transition from the superfluid to the Mott side dynamically. 
Either one decreases the tunneling rate in time or, alternatively, the
interaction is increased to suppress the superfluid density at given
filling and thus cross the transition line.  
It is demonstrated that in several particular cases for $J=J(t)$ or 
$U=U(t)$, respectively, the Bogoliubov-de~Gennes equations 
\eqref{Bogoliubov-deGennes} can be solved analytically. 

\subsection{Decreasing the tunneling rate}\label{decrease} 

Combining the two equations \eqref{dotdelta}, we get the following
second-order equation for the number fluctuations 
[analogously for phase fluctuations, see Eq.\,\eqref{EqmotionPhase} below]  
\bea
\left(
\frac{\partial}{\partial t}\,
\frac{1}{J }\,\frac{\partial}{\partial t}
+\lambda_\kappa
\left[J \lambda_\kappa +2 U n \right]
\right)
\delta\hat n_\kappa
=0
\,.\label{normalmodeEq}
\ea
Dividing this equation by $J$ and $\lambda_\kappa^2$, 
and defining a new time coordinate depending on the mode index
$\kappa$,
\bea 
d\tau_\kappa = \lambda_\kappa J dt \,,
\label{deftaukappa}
\ea 
we obtain an equation containing the operator 
$\partial^2/\partial\tau_\kappa^2$,  
\bea
\left(
\frac{\partial^2}{\partial\tau_\kappa^2}
+
1 + \frac UJ \frac{2 n}{\lambda_\kappa} 
\right)
\delta\hat n_\kappa
=0\,.
\label{taueom}
\ea
We note that this equation now contains the ratio $U/J$ only, but not 
$U$ and $J$ separately, cf.~Sec.~\ref{Increasing} below. 


Let us first study the case $J=J(t)$ while $U=\rm const$. 
Even though it is not possible to give a closed solution of
Eq.\,\eqref{normalmodeEq} for arbitrary time-dependences $J(t)$, there
are several cases which do admit analytic expressions in terms of 
Hankel $H_\nu^{(1)}$ or Whittaker $W_{\nu,\mu}$ functions
\cite{Abramowitz}.
We listed a few cases in table \ref{Jtable} below. 

In view of the asymptotic $t\uparrow\infty$ behaviour of the Hankel
and Whittaker functions \cite{Abramowitz}, we see that the number
fluctuations $\delta\hat n_\kappa$ oscillate forever in the last 
three cases ($J(t)\propto t^{-1},t^{-2/3},t^{-1/2}$) though with a
decreasing amplitude and frequency. 
In the case of an exponential -- i.e., much faster -- sweep 
$J(t)=J_0 e^{-\gamma t}$, on the
other hand, the solutions $\delta\hat n_\kappa$ do not have enough
time to adjust to the externally imposed change of $J(t)$ and freeze
at a finite value (non-adiabatic behaviour). 
Finally, the second case $J(t)=J_0 (\gamma t)^{-2}$ just marks the
border between the two regimes (eternal oscillation versus freezing).  
Consequently, the number fluctuations $\delta\hat n_\kappa$ vanish for
late times in this situation (for $\lambda_\kappa>0$). 

The asymptotic behaviour can be interpreted nicely in terms of the
analogy to cosmology sketched in Sec.~\ref{spacetime} below. 
The freezing of the number fluctuations $\delta\hat n_\kappa$ for
$J(t)=J_0 e^{-\gamma t}$ can then be understood via the emergence of a
horizon analogue -- whereas for 
$J(t)\propto t^{-1},t^{-2/3},t^{-1/2}$, such a horizon is absent. 
The critical behaviour $J(t)=J_0 (\gamma t)^{-2}$ precisely marks the
limit for horizon formation, cf.~Sec.~\ref{spacetime}. 

\begin{widetext}\noindent
\begin{table}[h]
\begin{tabular}{|c|c|c|c|c|}
\hline
$J(t)$
        & Solution
        & Argument 
        & Indices and constants
        & Asymptotics $t\uparrow\infty$
\\
\hline
$J_0 e^{-\gamma t}$
        & $W_{i\nu,\mu}(2ix_\kappa)$
        & $x_\kappa = \tau_\kappa = - {J_0\lambda_\kappa}e^{-\gamma t}/\gamma$
        & $\mu = {1}/{2}$, $\nu = {Un}/{\gamma}$
        & $x_\kappa\uparrow0$
\\
&&&&\\
$J_0 (\gamma t)^{-2}$
        & $\sqrt{x_\kappa} H_\nu^{(1)} ( x_\kappa )$
        & $x_\kappa = \tau_\kappa = - J_0 \lambda_\kappa(\gamma t)^{-1}/\gamma$
        & $\nu = \sqrt{1/4 - 2 UnJ_0\lambda_\kappa/\gamma^2}$
        & $x_\kappa\uparrow0$
\\
&&&&\\
$J_0 (\gamma t)^{-1}$
        & $H_\nu^{(1)} ( x_\kappa )$
        & $x_\kappa = 2\sqrt{2 U n J_0 \lambda_\kappa} 
(\gamma t)^{1/2}/\gamma$
        & $\nu = \pm 2i {J_0\lambda_\kappa}/{\gamma}$
        & $x_\kappa\uparrow\infty$
\\

&&&&\\
$J_0 (\gamma t)^{-2/3}$
        & $x_\kappa^{-1/4} W_{i\nu,\mu}( i x_\kappa )$
        & $x_\kappa = 3
                \sqrt{2 U n J_0 \lambda_\kappa} (\gamma t)^{2/3}/\gamma$
        & $\mu^2 = 1/16$,
         $\nu^2 = 9 (J_0\lambda_\kappa)^3/(32 U n \gamma^2)$
        & $x_\kappa\uparrow\infty$
\\
&&&&\\
$J_0 (\gamma t)^{-1/2}$
        & $\sqrt{x_\kappa} H_{\nu}^{(1)}
          ( c_\kappa x_\kappa^{3/2})$
        & $x_\kappa = \sqrt{\gamma t} + {J_0 \lambda_\kappa}/(2Un)$
        & $\nu=1/3,\,c_\kappa = \sqrt{2 UnJ_0\lambda_\kappa}/(3\gamma)$
        & $x_\kappa\uparrow\infty$

\\
\hline
\end{tabular}
\caption{\label{Jtable} A few analytically solvable cases 
for time-dependent $J=J(t)$. The last column indicates the behavior of the
argument $x_\kappa$ at late laboratory time.} 
\end{table} 
\noindent\vspace{.2cm}\end{widetext}

It is very illustrative to compare the various cases discussed above
to a harmonic oscillator with a time-dependent damping term and/or
spring constant. 
In the last three cases ($J(t)\propto t^{-1},t^{-2/3},t^{-1/2}$), the
spring constant dominates (underdamped oscillator) while the
exponential sweep $J(t)=J_0 e^{-\gamma t}$ induces a transition to the
overdamped regime at some time. 
The boundary case $J(t)=J_0 (\gamma t)^{-2}$ would then correspond to
critical damping, where the solution $\delta\hat n_\kappa$ approaches
zero very quickly. 
For the Bose-Hubbard model, this time-dependence would allow us to
approach the Mott state very efficiently. 

However, in an experimental realization, a dynamics 
$J(t)=J_0 (\gamma t)^{-2}$ requires some fine-tuning and is
probably hard to achieve. 
The exponential sweep is more interesting from a theory point of view
(since it yields non-zero frozen number fluctuations) and should also
be a better approximation to a realistic experimental situation. 
The experimental relevance becomes apparent considering the fact that, 
in sufficiently deep $d$-dimensional simple cubic lattices,
$J(t)\propto (V_0(t)/E_R)^{3/4} \exp\{-2\sqrt{V_0(t)/E_R}\}$
and $U \propto (V_0/E_R)^{d/4}$ hold, where
$V_0(t)$ is the time-dependent lattice depth given by the laser
intensity and $E_R=\pi^2/(2ma^2)$  
is the (constant) recoil energy, with $m$ the bare boson mass \cite{Boers}. 
Apart from logarithmically slow corrections, an exponential sweep
$J(t)\propto e^{-\gamma t}$ therefore corresponds to increasing
the laser amplitude (and therefore $\sqrt{V_0}$) linearly in time, 
with $U$ then remaining approximately constant.   

For the case of an exponentially decreasing tunneling rate, 
a universal ``scaling'' solution exists: 
As we may infer from the table \ref{Jtable}, the solution (i.e., the indices
$\mu$ and $\nu$ of the Whittaker function $W_{\nu,\mu}$) then do not depend
on $\kappa$. 
The ``scaling time'' from \eqref{deftaukappa}, depending on the mode index
$\kappa$, reads $\tau_\kappa=-\lambda_\kappa J(t)/\gamma $, so that 
Eq.\,\eqref{normalmodeEq} is transformed into a  
{\em scale invariant} equation of the form of Eq.\,\eqref{taueom}, 
\bea
\left(
\frac{\partial^2}{\partial\tau_\kappa^2}
+\left[1-\frac{2}{\tau_\kappa}\,\frac{Un}{\gamma}\right]
\right)
\delta\hat n_\kappa
=0
\,.\label{scalingEq}
\ea
The only remaining dimensionless parameter determining the relevant 
universality class of solutions of this equation is
$\nu=Un/\gamma$. 
This parameter, equal to the ratio of chemical potential $\mu=Un$ 
(i.e., internal energy scale) and sweep rate $\gamma$ 
(i.e., external time scale), represents a measure of the rapidity of
the externally imposed sweep: $\nu\gg1$ implies a slow and $\nu\ll1$ a
fast (non-adiabatic) sweep.  
The adiabaticity parameter $\nu$ determines the nature of the final
state. 
Starting in the superfluid phase and ramping down the tunneling rate
very rapidly, $\nu\ll1$, the system will have no time to react to this
change.
For later times, the vanishing hopping $J=0$ prevents an equilibration
of the number fluctuations, i.e., they will be as large as in the
initial coherent state in this situation. 
The slower we sweep $J(t)$, the more time the system has to adapt to
this external change and the closer the final state will be to the
Mott state (i.e., smaller number fluctuations will result from a slow
sweep).  

As listed in table \ref{Jtable}, 
the analytical solution of \eqref{scalingEq} can be found in terms of
the Whittaker functions $W_{i\nu,1/2}$ \cite{Abramowitz}. 
It allows us to obtain the following exact Bogoliubov transformation 
from bare density/filling fluctuation to initial vacuum quasiparticle
operators $\hat b_\kappa$, labeled by the mode index $\kappa$,  
\bea
\delta\hat n_\kappa =
\sqrt{n}\,e^{-\pi\nu/2}\,W_{i\nu,1/2}(2i\tau_\kappa)\,\hat b_\kappa
+{\rm h.c.} \label{analsolution} 
\ea
The initial vacuum quasiparticle operators $\hat b_\kappa$ annihilate 
the adiabatic superfluid ground state $\hat b_\kappa\ket{\rm in}=0$ 
at early times $\tau_\kappa\downarrow-\infty$, where the modes 
oscillate like $e^{\pm i\tau_\kappa}$. 
(Note that the scaling time $\tau_\kappa$ is dimensionless).
The phase fluctuations are obtained using the relation 
$\delta\hat\phi_\kappa =
-\frac{1}{2n} d \delta\hat n_\kappa/d\tau_\kappa$
following from \eqref{dotdelta} and \eqref{deftaukappa},  
\bea 
\delta\hat\phi_\kappa 
& = & -\frac{e^{-\pi\nu/2}}{2\sqrt{n}} \, 
\frac{dW_{i\nu,1/2}(2i\tau_\kappa)}{d\tau_\kappa}\,\hat b_\kappa
+{\rm h.c.} \label{phasefluctII}
\ea 
The analytical scaling solution thus obtained represents, to the  
best of our knowledge, the first example of an exact solution of the   
(nonintegrable) Bose-Hubbard model on the Bogoliubov mean-field level 
in a {\em dynamical} situation. 

Due to the perfect scaling solution in Eq.\,\eqref{analsolution}, 
the frozen value of the number fluctuations at late times, 
when $\tau_\kappa \rightarrow 0$, is independent of $\kappa$, but the 
decaying corrections do depend on the eigenvalue $\lambda_\kappa$:
\bea
\langle\delta\hat n_\kappa^2\rangle
\equiv
\bra{{\rm in}} (\delta\hat n_\kappa)^2 \ket{{\rm in}}
=
n\frac{1-e^{-2\pi\nu}}{2\pi\nu}
+\ord(t\lambda_\kappa e^{-\gamma t})
\,. \label{numberfluct} 
\ea
Since the leading term is independent of $\kappa$, it just yields a
local ($\propto\delta_{\alpha,\beta}$) contribution after the mode sum 
($\kappa\to\alpha$) and thus leads to frozen on-site number
variations. 

In contrast to the number fluctuations which freeze according to 
\eqref{numberfluct}, the conjugate phase fluctuations grow 
(as one would expect when approaching the Mott phase).
From our analytical result \eqref{phasefluctII}, we conclude that they 
increase (initially) quadratically in time: 
\bea
\langle\delta\hat\phi_\kappa^2\rangle
=
\nu\frac{1-e^{-2\pi\nu}}{2\pi n}\,\gamma^2t^2
+\ord(\gamma t\ln\lambda_\kappa)
\,. \label{phasefluct} 
\ea
Again, like for the number fluctuations, the 
leading (first) term is independent of the mode index $\kappa$ and  
yields the on-site phase fluctuations
$\langle\delta\hat\phi_\alpha^2\rangle$.  
The off-site phase correlations 
$\langle\delta\hat\phi_\alpha\delta\hat\phi_\beta\rangle$,
corresponding to the second term, grow linearly in time (initially).

\subsection{Final state}

All the results so far were obtained by a controlled
extrapolation of the mean-field expansion \eqref{mean-field} 
from the weak-coupling ($J$ dominates) into the strong-coupling regime  
($U$ dominates) and hence are only valid as long as the quantum
depletion $\langle\hat\chi^\dagger_\alpha\hat\chi_\alpha\rangle$ 
is small, i.e., the condensate $\psi_0$ dominates.  
In the strong-coupling regime, however, the quantum depletion grows
(on a time-scale of order $\sqrt{n}$) and finally invalidates the
mean-field expansion \eqref{mean-field}.
Fortunately, we may also analytically describe the ensuing stages of
the quantum evolution, because the tunneling rate
$J(t\gg1/\gamma)\lll1$ is exponentially small and can be completely  
neglected. 
In this completely interaction-dominated  limit, the evolution of the
site operators can be approximated by 
$d\hat a_\alpha/dt=-iU\hat n_\alpha\hat a_\alpha$, 
which possesses the simple exponential solution 
\bea
\label{U-dominated}
\hat a_\alpha(t)=\exp\{ -iU\hat n_\alpha^0 t \} \hat a_\alpha^0 
\qquad (J \lll 1)\,.
\ea
Consequently, we may calculate the time evolution of correlation
functions throughout the dynamics (i.e., for all $t$) by using the
results of the mean-field expansion, which are valid for intermediate
times with $\gamma t\gg1$ and $Ut\sqrt{n}\ll1$, as initial conditions, 
and then switching to the above solution \eqref{U-dominated} for later
stages.  

A frequently used indicator for distinguishing the superfluid from the 
Mott phase is the off-diagonal long-range order (ODLRO) usually
associated to the first-order correlation function 
$\langle\hat a_\alpha^\dagger(t)\hat a_\beta(t)\rangle$.
Using the results above, we may derive an analytical expression for
the time-dependence of 
$\langle\hat a_\alpha^\dagger(t)\hat a_\beta(t)\rangle$
during and after the sweep.
From \eqref{U-dominated}, we obtain for the correlator 
\bea
\langle\hat a_\alpha^\dagger(t)\hat a_\beta(t)\rangle
=n\langle\exp\{iU(\hat n_\alpha-\hat n_\beta)t\}\rangle
+\ord(\sqrt{n})
\,.
\ea
On the other hand, the frozen first-order number fluctuations
$\delta\hat n_\alpha$ are then in a squeezed state which can 
(for $n\gg1$) be approximated by a (continuous) Gaussian
distribution. 
For a Gaussian variable $X$ with $\langle X\rangle=0$,
the exponential average yields
$\langle\exp\{iX\}\rangle=\exp\{-\langle X^2\rangle/2\}$
and hence we finally get
\bea
\label{order}
\langle\hat a_\alpha^\dagger(t)\hat a_\beta(t)\rangle
\approx
n
\exp\{-U^2t^2\Delta^2(n_\alpha)\}
\,,
\ea
where the on-site number variations are given by \eqref{numberfluct} 
\bea
\Delta^2(n_\alpha)
=
\langle\hat n_\alpha^2\rangle-
\langle\hat n_\alpha\rangle^2
=
\langle\delta\hat n_\alpha^2\rangle
=
n\frac{1-e^{-2\pi\nu}}{2\pi\nu}
\,.
\ea
The result \eqref{order} represents an extension to the present period
lattice geometry of the result obtained in \cite{Imamoglu}, where the
interaction-induced decay of coherence in a double-well trap was
studied.  
The Fourier transform of the first-order correlation function 
$\langle\hat a_\alpha^\dagger(t)\hat a_\beta(t)\rangle$
determines the corresponding momentum distribution function 
$g_1({\bm k})$.   
The decay of the off-diagonal long-range order (ODLRO) 
in (\ref{order}) thus directly corresponds to a temporal decrease of
the peak in $g_1({\bm k})$ at ${k}=0$, measurable in time-of-flight
experiments \cite{GreinerII,Gerbier}. 

In summary, the state obtained with $J(t\gg1/\gamma)\lll1$ (while
still  maintaining $U t\sqrt{n}\ll1$, i.e. at intermediate times),
which has frozen number fluctuations according to \eqref{numberfluct},
forms an appropriate initial many-body quantum state for the further
evolution beyond mean-field into the strongly correlated Mott phase. 

\subsection{Increasing the interaction coupling}\label{Increasing}

Another possibility to approach the Mott phase dynamically is to
increase $U$ in time.  
This can be experimentally realized using a time-dependent sweep
through a Feshbach resonance, which varies the $s$-wave scattering 
length $a_s$, and thus $U$ only, while keeping $V_0$ and thus $J$
fixed.  
Since Eq.\,\eqref{taueom} depends on the ratio $U/J$ only, increasing
$U$ is analogous to decreasing $J$. 
Therefore, we can establish an exact ``duality'' between the results
obtained for $J=J(t)$ and the present $U=U(t)$, i.e., every analytic
solution in table \ref{Jtable} of 
Sec.~\ref{decrease} corresponds to a dual expression for
$U=U(t)$ after incorporating the transformation $t\to\tau$ of the time
coordinate in Eq.\,\eqref{deftaukappa}. 
For a power-law dependence $U\propto t^{\alpha}$, Eq.\,\eqref{taueom}
possesses formally the same solution as for $J\propto\tau^{-\alpha}$. 
Transforming back to the laboratory time via Eq.\,\eqref{deftaukappa}, 
this corresponds to $J\propto t^{-\alpha/(\alpha+1)}$ while the two
limiting cases $\alpha=-1$ and $\alpha=\infty$ correspond to
exponential dependences, see Tab.~\ref{duality}. 
Due to the transformation $t\to\tau$ of the time coordinate in
Eq.\,\eqref{deftaukappa}, the solutions for $U(t)$ do not freeze in
terms of the laboratory time $t\to\infty$.

\begin{table} 
\begin{tabular}{|c|c|c|}
\hline
$J(t)$ & Dual $U(t)$ & Dual Range 
\\
\hline
$J_0 e^{-\gamma t}$ & $U_0 (\gamma t)^{-1}$ & $-\infty<t<0$
\\
$J_0 (\gamma t)^{-2}$ & $U_0 (\gamma t)^{-2}$ & $-\infty<t<0$
\\
$J_0 (\gamma t)^{-1}$ & $U_0 e^{\gamma t}$ & $-\infty<t<\infty$
\\
$J_0 (\gamma t)^{-2/3}$ & $U_0 (\gamma t)^2$  & $0<t<\infty$
\\
$J_0 (\gamma t)^{-1/2}$ & $U_0 \gamma t$ & $0<t<\infty$
\\
\hline
\end{tabular}
\caption{\label{duality} Important dual cases for temporal
  variations of $U$ and $J$. Given the solution for the first column, we can
  immediately conclude, by a simple time transformation, on the
  solution for the second column, and vice versa.}
\end{table}

Again, it is likely that 
most of the dynamics $U(t)$ are hard to realize
experimentally. 
Therefore, we focus on the linear increase, $U(t)=\gamma t$, in the
following, since this case is probably close to an experimental
set-up.  
For a linear growth (dually corresponding to $J(t)\propto1/\sqrt t$),  
we can define a scaling time using a simple $\kappa$-dependent shift
of the time origin, 
\bea
\tau_\kappa=t+\frac{J\lambda_\kappa}{2n\gamma} \quad
\Leftrightarrow\quad 
d\tau_\kappa = dt \,,
\ea
Introducing this scaling time into the equation for phase 
fluctuations 
\bea
\left(
\frac{\partial}{\partial t}\,
\frac{1}{2nU(t)+J\lambda_\kappa}\,
\frac{\partial}{\partial t}
+J\lambda_\kappa
\right)
\delta\hat\phi_\kappa
=0
\,, \label{EqmotionPhase} 
\ea
we obtain a 
scaling equation of the following form 
\bea
\left(
\frac{\partial}{\partial\tau_\kappa}\,
\frac{1}{\tau_\kappa}\,
\frac{\partial}{\partial\tau_\kappa}
+2n\gamma J\lambda_\kappa
\right)
\delta\hat\phi_\kappa
=0
\,.
\ea
The solution of this equation can be found in terms of Hankel
functions and leads us to the following Bogoliubov transformation for
the phase fluctuations 
\bea
\delta\hat\phi_\kappa
=
\hat C_\kappa \tau_\kappa\,H^{(1)}_{2/3}
\left(\frac{2}{3}\,\sqrt{2n \gamma J\lambda_\kappa}\,
\tau_\kappa^{3/2}\right)
+{\rm h.c.}
\label{Uanalsolution} 
\ea
where $H^{(1)}_\nu$ and $H^{(2)}_\nu = (H^{(1)}_\nu)^*$ are Hankel
functions of the first and second kind, respectively. Note that the 
index is mode number $\kappa$-independent, like the solution for 
exponential decrease of $J$ in \eqref{analsolution}. 
In contrast to the above-discussed case of an exponential decay of
$J$, approaching the ``hard-core'' limit by letting $U$ grow
(linearly), we do not obtain frozen number fluctuations at late 
times $\tau_\kappa\uparrow\infty$. 
We then have, instead, increasingly rapid oscillations of fluctuating
filling 
\bea
\delta\hat n_\kappa
&=& - \frac1{\gamma\tau_\kappa} \frac{d \delta \phi_\kappa}{d\tau_\kappa} 
\nn 
& = & \hat {\tilde C}_\kappa \tau_\kappa^{1/2}\,H^{(1)}_{1/3}
\left(\frac{2}{3}\,\sqrt{2n\gamma J\lambda_\kappa}\,\tau_\kappa^{3/2}\right) 
+{\rm h.c.}
, \label{nkappaU(t)}
\ea
where the operators $\hat {\tilde C}_\kappa$ are proportional to the 
$\hat C_\kappa$ in \eqref{Uanalsolution}.
Due to the asymptotic (large $x$) behavior 
$H_\nu(x) \rightarrow 
\sqrt{2/{\pi x}}\exp[i(x-\nu \pi /2 -\pi/4)]$, we have at late times 
$\delta\hat n_\kappa \propto t^{-1/4} 
\exp[i\frac{2}{3}\,\sqrt{2n\gamma J\lambda_\kappa} t^{3/2}]+$h.c.  
From \eqref{Uanalsolution}, we derive the corresponding slow increase
of the amplitude of phase fluctuations, 
$\delta\hat\phi_\kappa \propto t ^{1/4} 
\exp[i\frac{2}{3}\,\sqrt{2n\gamma J\lambda_\kappa} t^{3/2}]+$ h.c., 
with the same increase of the oscillation frequency.

\subsection{Quantum depletion} 

Approaching the Mott phase entails that the condensate becomes
depleted due to Bogoliubov quasiparticle excitations created above the
(still superfluid) ground state. 
We stress that the quantum depletion we calculate below explicitly
depends on $J=J(t)$ [or $U=U(t)$] in a particular manner. 
Thus it is possible, within our controlled mean-field scheme, to 
unambiguously identify the number squeezing (related to the depletion) 
caused by the creation of the excitations, and to distinguish it from
effects coming from the time dependence of the mean-field in the
Gross-Pitaevski\v\i\/ equation (which were discussed in \cite{McKagan}). 

The (relative) depletion is defined by the expression 
\bea 
{\cal D} \equiv 
\frac{\sum_\alpha\langle \hat \chi^\dagger_\alpha\hat \chi_\alpha 
\rangle} {\sum_\alpha n_\alpha}
\,. 
\ea
Concentrating here on $J=J_0\exp [-\gamma t]$, where as shown in
section \ref{decrease} number fluctuations freeze, and using that the
local depletion is given by 
$ \hat \chi^\dagger_\alpha\hat \chi_\alpha = 
\delta \hat n_\alpha^2/(4n) + n\delta \hat \phi_\alpha^2+1/2 $, 
we have from \eqref{numberfluct} and \eqref{phasefluct}, 
\bea 
{\cal D} &=&
\frac{\langle \hat \chi^\dagger_\alpha\hat \chi_\alpha 
\rangle} {n}
\simeq  \frac{1}{2\pi n} \left(\nu \gamma^2 t^2 +\frac1\nu\right)
\left(1-e^{-2\pi\nu}\right)+ \frac1{2n} \nn
& \simeq & \begin{cases}
\displaystyle U^2 n t^2 
& \nu \ll 1 \smallskip\\
\displaystyle 
\frac{U\gamma t^2}{2\pi} 
& \nu \gg 1 
\end{cases} 
\ea
which is valid provided that the mean-field condition $Ut\sqrt{n}\ll1$
holds -- which implies ${\cal D}\ll1$. 
The phase fluctuations caused by the sweep always dominate the number
fluctuations in the limit assumed throughout, $n\gg 1$ 
(while still keeping $Ut\sqrt{n}\ll 1 $).  

\subsection{Comparison to continuum scaling} 

Up to now we have treated the homogeneous case. 
If the lattice is embedded in an external harmonic trapping potential, 
like in experiment, the derived analytical solutions for 
$J\propto\exp[-\gamma t]$ and $U\propto t$ are not exactly valid
anymore.  
However, in the continuum limit, the well-known scaling approach for a
bulk gas in a harmonic trapping potential \cite{Kagan,Dum,Plata} can
be applied and it is rather interesting to contrast this approach with 
our scaling solutions on the lattice \eqref{analsolution} and
\eqref{nkappaU(t)}. 

The continuum scaling equations of motion in 1D read 
\bea
\ddot b(t) + \frac{J(t)}{J_0} \omega^2_0 b(t) -
\frac{\dot{J}(t)}{J(t)} \dot b(t) 
=  
\frac{J(t)}{J_0}\,
\frac{U(t)}{U_0}\,
\frac{\omega_0^2}{b^2(t)}, 
\label{scalingbz} 
\ea
where $b(t)$ denotes the time-dependent scaling parameter 
\cite{Kagan,Dum,Plata} while $\omega_0$, $J_0$, and $U_0$ are the 
initial values for trap frequency, tunneling, and interaction rate,
respectively.  
We conclude from Eq.\,\eqref{scalingbz} that the temporal change of
$J(t)$ corresponds to a change in the (inverse) effective mass of the
bosons.   

For $J(t)=J_0 e^{-\gamma t}$, the effective harmonic trapping
frequency becomes exponentially slower,  
and the scaling factor motion consequently overdamped due to the
resulting constant (Ohmic) damping. 
This damping corresponds to the derived freezing of the number
fluctuations and increase of phase fluctuations on the lattice:  
\bea
\ddot b(t) + \gamma \dot b(t) + e^{-\gamma t}\omega_0^2 b(t) =   
\frac{e^{-\gamma t}\omega_0^2}{b^2(t)} 
\label{damping} 
\,.
\ea
Conversely, for $U(t)$ linearly increasing in time, and constant $J$,
due to the driving term on the right-hand side of \eqref{scalingbz}, 
the scaling factor
begins to oscillate increasingly fast, like already observed from  
the exact solution on the lattice, Eq. \eqref{nkappaU(t)}. 

\section{Nonequilibrium Quantum Effects in Inhomogeneous Systems}

In the additional presence of an optical lattice, the behavior of the
system depends on the relation between the inhomogeneity of the trap
potential $V_\alpha$ and the central filling $n_{\rm center}\gg1$. 
If the potential $V_\alpha$ is rather shallow, the system develops a
``wedding cake'' structure near the boundaries, where the filling is
small and hence the Mott phase emerges \cite{BosonicMott,Foelling06}.
For stronger inhomogeneities of $V_\alpha$ and/or larger central
fillings $n_{\rm center}\gg1$, on the other hand, two-body
interactions $U(\hat a_\alpha^\dagger)^2\hat a_\alpha^2$ in 
\eqref{Hamiltonian} can be neglected in comparison with the tunneling
term $JM_{\alpha\beta}\hat a_\alpha^\dagger\hat a_\beta$ and the
potential gradient of $V_\alpha \hat n_\alpha$ due to  
$U=\ord(1/n_{\rm center})$.
Intuitively speaking, the various rims of the ``wedding cake'' would
be compressed into a single lattice site and hence disappear. 

Lowering $J$ towards the (first) phase transition point, the wedding
cake structure gradually develops by Mott insulator shells propagating
inwards from the outer low density edges of the gas cloud. 
Therefore, we have to study the question of whether the results
derived above for the homogeneous case -- such as the temporal decay
of the coherence peak of the momentum distribution function 
$g_1({\bm k})$ due to number squeezing effects -- can still be
observed in the presence of an external trap.  
Clearly, near the boundaries inhomogeneity effects will be important.
However, in the central region, the homogeneity assumption should
provide a good approximation -- provided that the 
momentum distribution function $g_1({\bm k})$ 
decays fast enough, i.e., before the Mott insulator
shells propagating inwards reach the center. 
From Eq.\,\eqref{order}, we conclude that the  typical time for the
peak at zero wave vector of the momentum distribution function  
$g_1({k=0})$ to decay
(say, one $e$-folding) is generally scaling with $1/\sqrt n$, i.e., we have
\bea
t_d = \frac1{U\sqrt n} \quad (\nu\ll 1), \qquad \quad 
 t_d = \frac1{U}\sqrt{\frac\nu n} \quad (\nu\gg 1)
\,.  
\ea 
To clarify the question above without resorting to brute force numerical
computation, using simple analytical means, we consider in what
follows various possible modes of the disturbance propagation of the
corrections to mean-field, to determine the dominant, i.e. most rapid,
mode of propagation. 

\subsection{Propagation of sound and shock waves}

The speed of sound in a 1D optical lattice, for large filling $n$, 
can be derived from the dispersion law \cite{Javanainen} 
\bea
\omega_k^2 = 4n U J \sin^2 \left(\frac{ka}2\right) + 
4J^2 \sin^4 \left(\frac{ka}2\right), \label{dispersion}
\ea 
where the lattice spacing $a=\lambda/2$ is determined by the laser
wavelength $\lambda$ 
[the eigenvalues of $M_{\alpha\beta}$ in this case are 
$\lambda_k= 2\sin^2 (ka/2)$, from which, using \eqref{spectrum},
the above dispersion law follows]. 
This leads for $k\rightarrow 0$ to the sound speed  
\bea
c_s = a \sqrt{nUJ} = \pi \sqrt{\frac{nUJ}{2mE_R}} 
\,.  
\label{sounddispersion} 
\ea
A condition for the undisturbed observability of the decay
of $g_1 (k)$ is, then, that the time for sound perturbations to propagate
to the center is much larger than the decay time $t_d$, i.e., 
$R_{TF}/(c_s)_0\gg t_d$, where the Thomas-Fermi size of the
harmonically trapped cloud $R_{TF}=\sqrt{2mUn_0/\omega^2}$,  
with $n_0$ the central filling and central sound speed  
$(c_s)_0 = c_s (J=J_0, n=n_0)$.
We then obtain that, using $J_c\sim U/n$ 
\bea 
\frac {2U}{\pi \omega} \sqrt{\frac{E_R}{J}} \gg \frac1{\sqrt{n}}
\label{inequality}\quad  
\Longleftrightarrow  \quad 
 \frac {2}{\pi \omega} \sqrt{\frac{U E_R}{J/J_c}} \gg \frac1{n}
\ea 
has to be fulfilled for a rapid quench, $\nu \ll 1$.
This condition is practically always fulfilled if the optical lattice
is not too tightly harmonically trapped: 
With a laser wavelength of $\lambda =985$\,nm, we have 
$E_R\simeq 3.7$\,kHz for $^{87}\!$Rb and $E_R\simeq 8.9$\,kHz for Na, 
while typical values for the on-site interaction are 
$U\sim 10^{-2} \cdots 10^{-3}$\,$E_R$ \cite{Morsch}; furthermore,   
$\omega$ in the weakly confining direction(s) does not exceed 100\,Hz.
For a slow quench, the right-hand side of the above inequality
\eqref{inequality} reads $\sqrt{\nu/n}$ (respectively $\sqrt{\nu}/n$) 
instead, and the inequality will only be 
fulfilled if $\nu$ is not too large, i.e., if the sweep is not too slow
(as one would expect). 

It is a well-established fact that shock waves in a homogeneous medium 
can propagate much faster than sound;
we therefore need to estimate if shock waves might overtake the sound
waves in the proceeding quench. 
However, according to the study of \cite{Kraemer}, the propagation of
shock waves in the {\em optical lattice} proceeds at speeds 
{\em slower} than that of the sound waves for all values of 
$ J < \frac13 n U $; the dispersion relation \eqref{dispersion} then
has negative curvature for all $k$. 
This condition translates into $(J/{J_c}) (1/n^2) \lesssim \frac 13$. 
Therefore, close to the Mott transition, and for sufficiently 
large filling, ``shock'' waves always propagate {slower} than
sound, as opposed to the uniform system.

\subsection{Propagation of the phase boundary between superfluid and
  Mott phases} 

\begin{figure}[hbt]
\centerline{
\mbox{\epsfxsize=3cm\epsffile{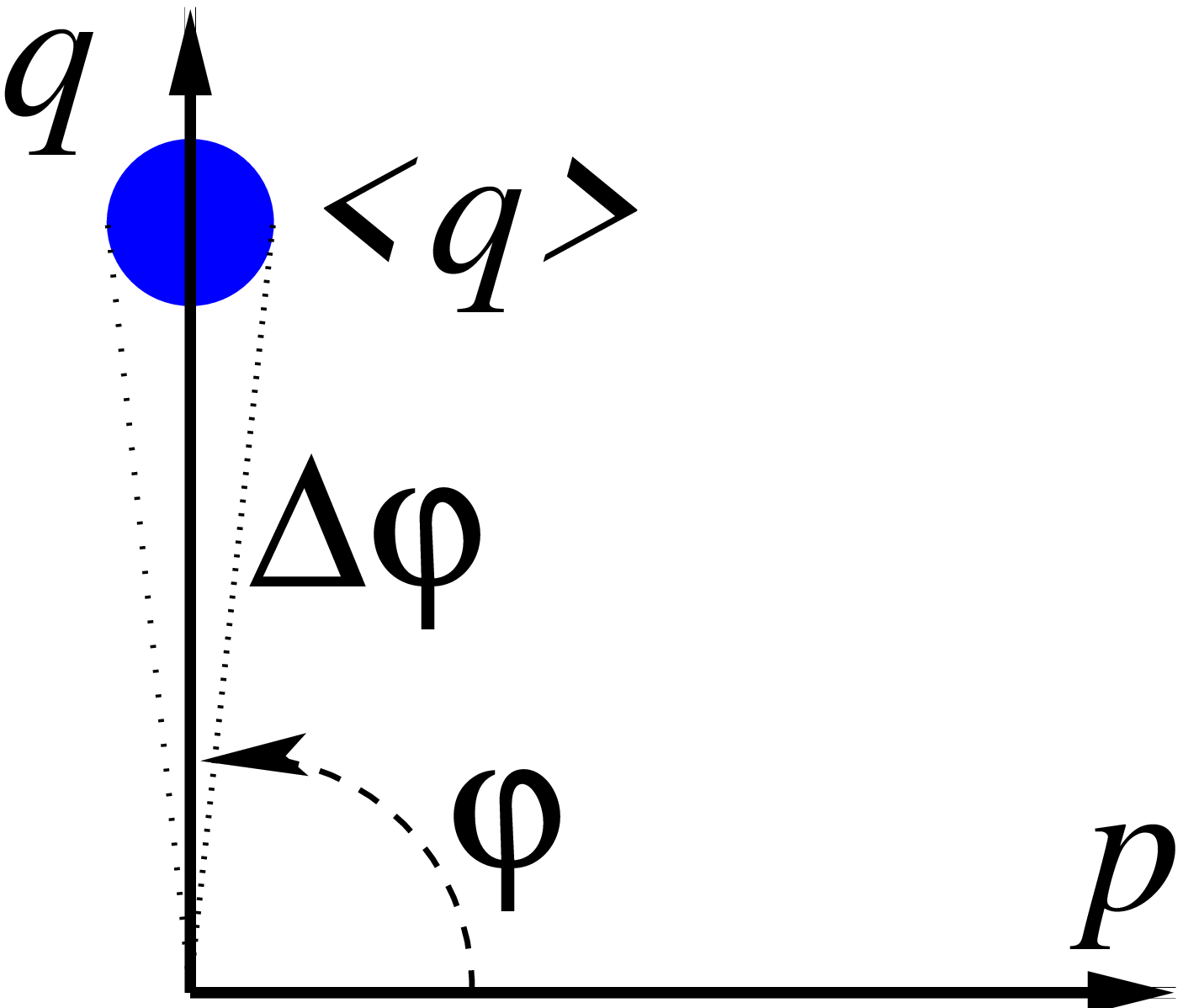}}
\hspace{1cm}
\mbox{\epsfxsize=3.5cm\epsffile{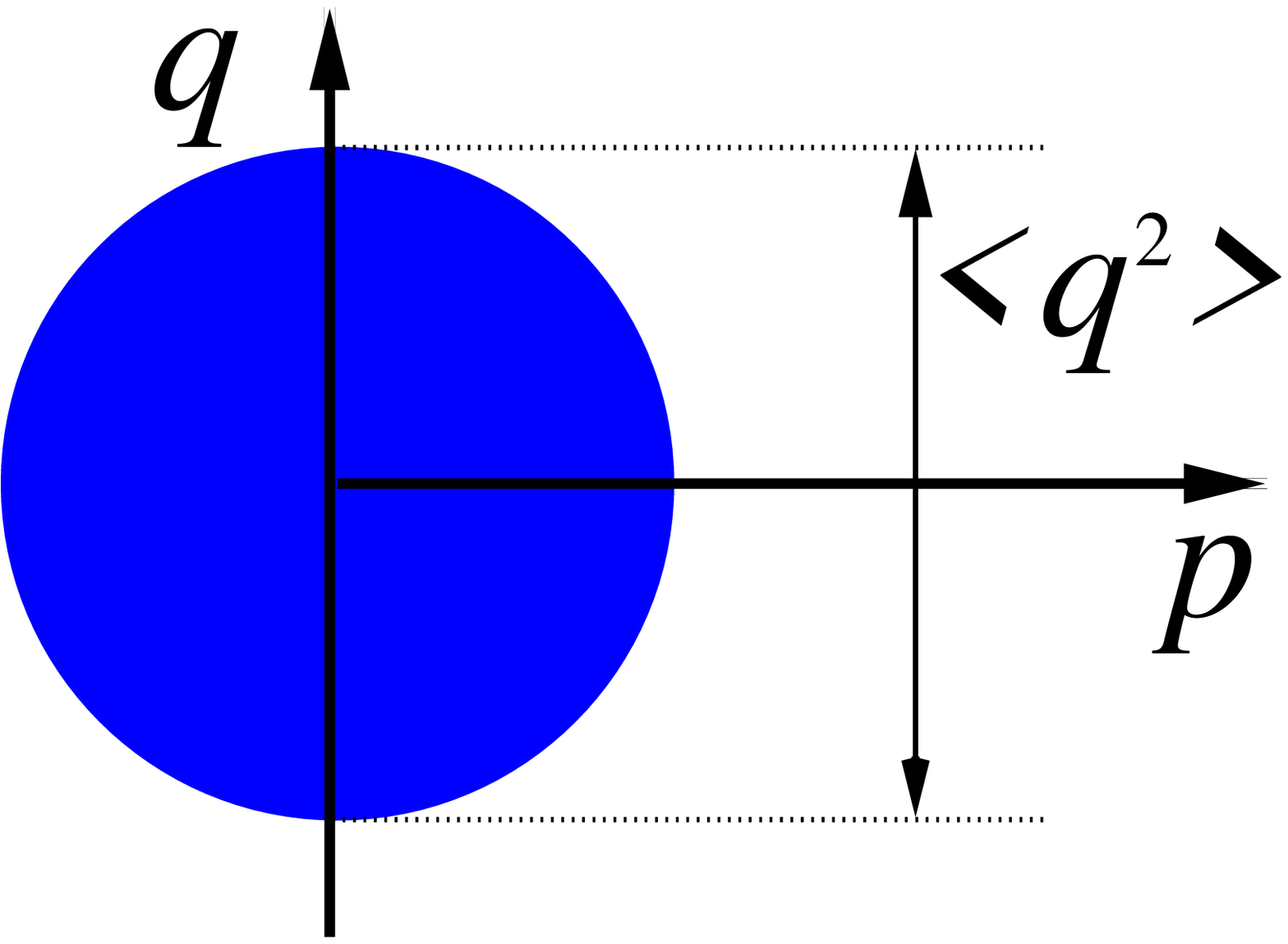}}
}
\caption{\label{phasespace} 
[Color online]
Phase space diagram (not to scale) of the coherent state
  (left) and the number eigenstate (right).}
\label{zeiger}
\end{figure}

In the two examples above (sound and shock waves), we studied the
propagation within a {\em given} phase (the superfluid phase).
However, one might object that the motion of the superfluid-Mott phase
{\em boundary} could perhaps be much faster.
Clearly, the typical time scale to overcome one lattice site will
again be set by the tunneling rate $J$, but this rate could possibly be
enhanced (or suppressed) by a large filling factor $n\gg1$.
In general, the correct description of the propagation of a phase
boundary in the Bose-Hubbard model is a very interesting and quite
involved problem.
In the following, we derive a rough estimate using the simpler two-site 
Bose-Hubbard model (a Bose-Einstein condensate Josephson junction),  
described by the Hamiltonian 
\bea
\hat H=J(\hat a_1\hat a_2^\dagger+\hat a_1^\dagger\hat a_2)+\ord(U)
\,.
\ea
In one site ($\hat a_1$), we model the superfluid phase by a coherent
state, while the other side ($\hat a_2$) is prepared in a number
eigenstate simulating the Mott phase, cf. Fig.\,\ref{phasespace}.
Therefore, the combined initial state factorizes 
\bea
\ket{\psi_{12}^0}=\ket{\psi_{1}^0}\otimes\ket{\psi_{2}^0}
\,,
\ea
into a coherent state $\ket{\psi_{1}^0}$ and a number eigenstate
$\ket{\psi_{2}^0}$ 
\bea
\hat a_1\ket{\psi_{1}^0}=\alpha_1\ket{\psi_{1}^0}
\;,\,
\hat n_2\ket{\psi_{2}^0}=n_2\ket{\psi_{2}^0}
\,.
\ea
Now the question is the following: how fast does the phase coherence
of the coherent state (superfluid) decay due to the coupling with the
other site (Mott)?
Since the propagating superfluid-Mott phase boundary is (like a shock
wave) presumably very sharp, this site-to-site behavior should
provide a rough stimate for the propagation speed.
Of course, there will also be an on-site dephasing caused by the
self-interaction $U$, but this process precisely corresponds to the
homogeneous growth of the phase fluctuations considered in the present 
Article and is independent of the moving phase boundary.
\bigskip

Without loss of generality, we assume $\alpha_1\in\mathbb R$, i.e.,
$\langle\hat p_1\rangle_0=0$ and $\langle\hat q_1\rangle_0\neq0$ in
terms of the canonical variables 
$\hat q_1=(\hat a_1^\dagger+\hat a_1)/\sqrt{2}$ and
$\hat p_1=i(\hat a_1^\dagger-\hat a_1)/\sqrt{2}$. 
The initial phase uncertainty is given by, cf. Fig.\,\ref{phasespace} 
\bea
\langle(\Delta\hat\varphi_1)^2\rangle_0
=
\frac{\langle\hat p_1^2\rangle_0}{\langle\hat q_1\rangle_0^2}
=
\ord\left(\frac{1}{\alpha_1^2}\right)
=
\ord\left(\frac{1}{\langle\hat n_1\rangle}\right)
\,,
\ea
which is the usual result for a coherent state. 
Now, if we insert the time evolution in the Heisenberg picture  
\bea
\hat p_1(t)=\hat p_1(0)+Jt\hat q_2(0)+\ord(J^2t^2)
\,,
\ea
we find a gradual increase of the phase variance 
\bea
\langle(\Delta\hat\varphi_1)^2\rangle(t)
=
\langle(\Delta\hat\varphi_1)^2\rangle_0
+J^2t^2
\frac{\langle\hat q_2^2\rangle_0}{\langle\hat q_1\rangle_0^2}
\,,
\ea
which just corresponds to the dephasing induced by the coupling to the
other site. 
Since the phase fluctuations of the other site are maximally large 
$\langle\hat q_2^2\rangle_0=\ord(n_2)$, we get together with 
$\langle\hat q_1\rangle_0=\ord(\sqrt{n_1})$ the final result 
\bea
\langle(\Delta\hat\varphi_1)^2\rangle(t)
=
\ord\left(\frac{1}{n_1}\right)+
\ord\left(J^2t^2\frac{n_2}{n_1}\right)
\,.
\ea
Hence, for equal (mean) fillings $n_1=n_2=n$, the typical dephasing
time scale $\langle(\Delta\hat\varphi_1)^2\rangle=\ord(1)$ is just
determined by the tunneling rate $J$ without suppression or
enhancement by additional factors of $n\gg1$. 

To summarize, the propagation velocity of the two-site model for the
phase boundary is of order $v_2=\ord(aJ)$.
The ratio of the maximal sound speed from \eqref{sounddispersion} 
to $v_2$ is given by
\bea 
\frac{c_s}{v_2} \sim \sqrt{\frac{nU}J} \sim \frac{n}{\sqrt{J/J_c}}
\,.  
\ea
It follows that sound waves are the dominant, i.e., most rapid mode of
propagation of the disturbances caused by the Mott insulator shells,
due to the $n$-dependent collective enhancement factor, which is
absent at least from the simple two-site model for the phase boundary
propagation.  

\section{Effective spacetime at large wavelengths}\label{spacetime}

The freezing of the number fluctuations obtained in section
\ref{decrease} can nicely be explained via the analogy to the
kinematics of quantum fields in gravity, valid at large length scales
and low energies.  
At large wavelengths $\lambda \gg a$, the lattice structure is not
important for the propagation of disturbances and the system behaves
like an ordinary (possibly inhomogeneous and moving) superfluid. 
[This is precisely the limit in which the scaling
  Eq.\,\eqref{scalingbz} holds.]  
In the continuum limit, the excitations (phonons) possess a linear
spectrum at low energies 
and behave in complete analogy to scalar (quantum) fields in
curved space-times \cite{Unruh,Visser}.

The case of a decreasing tunneling rate $J(t)$ is analogous to an
expanding universe. 
This can be seen by means of the evolution equation
\eqref{normalmodeEq} in the limit of small wavenumbers 
$k=\sqrt{8\lambda_\kappa}/a$ 
\bea
\left(
\frac{\partial}{\partial t}\,
\frac{1}{J(t)}\,\frac{\partial}{\partial t}
-Una^2 \nabla^2 
\right)
\delta\hat n (x,t) 
=0
\,,
\label{UniverseEq}
\ea
which is identical to the wave equation for a scalar field mode in an 
expanding universe \cite{Inflation,Ralf}. 
If the speed of sound $c^2_s(t)=Una^2J(t)$ decreases fast enough
(e.g., exponentially), phonons cannot propagate arbitrarily far but
may only travel a finite distance. 
In analogy to cosmology, this corresponds to the emergence of a
horizon 
\bea
\Delta_{\rm h}(t) = \int\limits_t^\infty dt'\,c_s(t') = 
a\int\limits_t^\infty dt'\,\sqrt{J(t')Un} 
\,.
\label{horizon}
\ea
The convergence of the above integral for $t'\to\infty$ then indicates
the emergence of a horizon. 
This observation matches the conclusions of Section \ref{decrease},
where we found that the solutions for $J\propto t^{-\alpha}$ with
$\alpha<2$ do not freeze at late times but oscillate forever 
(no horizon).  
For an exponential sweep, on the other hand, the integral converges
and a horizon exists.  
Because its size shrinks exponentially 
$\Delta_{\rm h}(t)={2\sqrt{J_0 U n}}\exp[-\gamma t/2]/\gamma$, it  
engulfs any given mode with wavelength $\lambda$ after some time.  
%
%
Therefore, the evolution of the phonon modes passes through three
stages: oscillation $\lambda\ll\Delta_{\rm h}(t_<)$, horizon crossing 
$\lambda=\Delta_{\rm h}(t_=)$, and freezing 
$\lambda\gg\Delta_{\rm h}(t_>)$. 
Since each mode can be mapped to a harmonic oscillator, this evolution
corresponds to the transition from underdamped to overdamped regime,
which can be seen by rewriting \eqref{UniverseEq} as  
\bea
\left(
\frac{\partial^2}{\partial t^2}
-
\frac{\dot J}{J}\,\frac{\partial}{\partial t}
-
J(t)Una^2 \nabla^2 
\right)
\delta\hat n (x,t) 
=0
\,.
\label{oscillator}
\ea
Hence, identifying $\dot J/J$ with the Hubble constant (which is indeed
a constant for the exponential sweep), 
the freezing of the number fluctuations and the creation of a
number squeezed state by the exponential 
sweep considered in this Article is completely
analogous to cosmic inflation \cite{Inflation}.
During this very early epoch of the cosmic evolution, the rapid
expansion of space induced a squeezing of the quantum fluctuations of 
the inflaton scalar field -- traces of these frozen and amplified
quantum fluctuations can still be observed today in the anisotropies
of the cosmic micro-wave background radiation. 


\section{Conclusions}
 
For the Bose-Hubbard model with generally 
time-dependent coefficients $J(t)$ and
$U(t)$, we developed a rigorously controlled and number-conserving
mean-field expansion for large filling $n\gg1$ as a generalization of
the original theory of Bogoliubov to the lattice. 
This allows us to study non-equlibrium quantum phenomena occuring in
the sweep from the superfluid to the Mott phase.
For two experimentally relevant cases -- an exponential decay of the
tunneling rate $J(t)$ and a linear increase of the interaction
coupling $U(t)$, respectively -- we found exact scaling solutions which
facilitate fully analytical expressions for the time-dependence of the 
Bogoliubov excitations and the resulting quantum depletion. Moreover,
we were able to establish a duality between various power-law behaviors
of $U(t)$ and $J(t)$, which enables transferring one solution obtained 
for $J=J(t), U=$\,const. into one for $U=U(t), J=$\,const., and vice
versa.

In the first case (exponentially decaying tunneling rate), we observe freezing
and squeezing of the initial quantum number fluctuations in complete 
analogy to cosmology.
The final state and its energetic position between the (initial)
coherent state and the Mott phase (final ground state) depends on  
a single adiabaticity parameter $\nu$, which is given by the ratio of
the (external) sweep rate and the (internal) chemical potential.
This parameter also governs the decay of off-diagonal long-range
order.  

Since the Bose-Hubbard model is considered a prototypical example for
quantum criticality, we expect our findings to contribute to the
general understanding of dynamical quantum phase transitions.
Furthermore, by estimating the applicability of our results derived
for the homogeneous case to the real experimental situation with 
harmonic trapping present, we found
that the predicted effects should be observable with current optical 
lattice technology.  

%

\section*{Acknowledgments} 

This work was supported by the Emmy Noether Programme of the
German Research Foundation (DFG) under grant No.~SCHU~1557/1-2,3
and by the Australian Research Council.


\begin{thebibliography}{9999}
\bibitem{BdGfootnote} 
We remark that a gapped amplitude mode emerges from the rotor
Hamiltonian \eqref{rotorH} near the phase transition if one uses a
dynamical Gutzwiller type mean-field ansatz for the ground state on
the lattice \cite{Huber}, for which the condensate will be strongly
depleted.   
%
Within Bogoliubov mean-field theory, i.e., using the weakly depleted
superfluid ground state, such a mode does not occur, and the
excitation spectrum 
(given by Eq.\,\eqref{spectrum} in the time-independent case)  
always remains gapless.
\end{thebibliography}
\end{document}